\documentclass[preprint,11pt,showpacs,preprintnumbers,amsmath,amssymb]{revtex4}
\usepackage{epsfig}
\usepackage{graphicx}
\usepackage{dcolumn}

\newcommand{\beq}{\begin{equation}} \newcommand{\eeq}{\end{equation}}
\newcommand{\beqa}{\begin{eqnarray}}
\newcommand{\eeqa}{\end{eqnarray}}

\begin{document}

\title{Quantum Brownian Motion of a Macroscopic Object in a General Environment}
\author{Chung-Hsien Chou$^{1,2}$\footnote{Email Address:
chouch@phys.sinica.edu.tw}, B.~L. Hu$^{2}$\footnote{Email address:
blhu@umd.edu} and Ting Yu$^{3}$\footnote{Email address:
ting@pas.rochester.edu}} \affiliation{$^1$Center for Gravitation,
Cosmology and Quantum Physics, Institute of Physics, Academia
Sinica,
Nankang, Taipei 11529, Taiwan\\
$^2$Joint Quantum Institute and Maryland Center for Fundamental
Physics, \\ Department of Physics, University of
Maryland, College Park, Maryland 20742-4111\\
$^3$Department of Physics and Astronomy, University of Rochester,
Rochester, New York 14627-0171}
\date{August 7, 2007}

\begin{abstract}
For the purpose of understanding the quantum behavior such as
quantum decoherence, fluctuations, dissipation, entanglement and
teleportation of a mesoscopic or macroscopic object interacting
with a general environment, we derive here a set of exact master
equations for the reduced density matrix of $N$  interacting harmonic
oscillators in a heat bath with arbitrary spectral density and
temperature.  Two classes of problems of interest to us which
these equations can be usefully applied to are that of the quantum
dynamics of nanoelectromechanical oscillators and the entanglement
evolution of multi-partite macroscopic states such as quantum
superposition of mirrors in a high Q cavity. To address a key
conceptual issue for macroscopic quantum phenomena we examine the
conditions for an assumption often implicitly made in these
studies to be valid, namely, that the quantum behavior of a
macroscopic object in an environment can be accurately represented
by only treating the dynamics of its center-of-mass variable. We
also mention how these results can be used to calculate the
uncertainty principle governing a macroscopic object at finite
temperature.
\end{abstract}


\maketitle

\section{INTRODUCTION}

In an earlier paper \cite{2HOQBM}  we showed the derivation of an
exact master equation for two coupled quantum harmonic oscillators
interacting bilinearly with a common environment made up of $n$
harmonic oscillators at an arbitrary temperature for a general
spectral density function. (This is referred to  as a `general
environment' in \cite{HPZ1}.) This equation can be applied to the
analysis of model problems in macroscopic quantum phenomena (MQP)
involving two harmonic oscillators, either mechanical such as the
superposition of two mirrors \cite{Marshall,Adler,Diosi}, or one of
them of electromagnetic or superconducting flux origin,  as in
nanoelectromechanical (NEM) resonators \cite{BlencoweRev} or
SQUID-resonator \cite{BuksBlen06} \footnote{A level reduction scheme
need be applied to one of the two oscillators, see, e.g.,
\cite{ShiHu04}, thus turning it into a two-level system. The
corresponding transformed equation can be applied to problems modeled
by a qubit-oscillator interaction (see, e.g., \cite{Bose06}).}.

In this paper we take a small step towards treating the quantum
properties of a mesoscopic or macroscopic object in a general
environment by providing the technical base for such studies. We
consider a system modeled by $N$ harmonic oscillators (NHO)
interacting with a heat bath consisting of $n$ harmonic oscillators
(HOB). The aim is to delineate the conditions upon which the
mechanical and statistical mechanical properties of this quantum
object can be described in terms of its center-of-mass (COM) variable
by a master equation for the reduced density matrix with the bath
variables integrated out, and for such conditions, derive an exact
master equation for a bath with arbitrary spectral density and
temperature. The motivation for this work has both conceptual and
practical underpinnings. At the conceptual level we want to examine
the validity of an implicit assumption made in many MQP
investigations, namely, that the quantum mechanical behavior of a
macroscopic object like the NEM or a C60 molecule
\cite{Arndt,Brezger}, placed in interaction with an environment,
behavior such as quantum decoherence \cite{BJK99}, fluctuations and
dissipation, entanglement and teleportation, can be captured by its
COM behavior. For convenience we refer to this as the `COM axiom'.
This assertion is intuitively reasonable, as one might expect it to
be true from normal- mode decompositions familiar in classical
mechanics, but when  particles (NHO) interact with each other (such
as in a quantum bound state problem) in addition to interacting with
their common environment, all expressed in terms of the reduced
density matrix, it is not such a clear-cut result. At least we have
not seen a proof of it \footnote{In realistic situations, just saying
that the object contains N particles is not enough; there are layers
of structure involved. Are the particles molecules, atoms, nucleons
or quarks? The coupling strength between constituents at each level
of structure (e.g., inter-atomic) compared to that structure's
coupling with the environment (e.g., atom- field in cavity QED, where
the field is taken to be the environment) will determine the relative
weight of each level of structure's partaking of the macroscopic
object's overall quantum behavior. It is for this reason that in
atom-optical (ev scale) physics we don't usually mention quarks of
the GeV energy scale. If we are interested in atomic scale processes
we may refer to atoms as physically `relevant' particles and all
sub-level particles as `irrelevant' (borrowing terminology from
projection operator formalism in nonequilibrium statistical
mechanics), as far as their contribution to the macroscopic quantum
phenomena at the atomic physics energy scale is concerned. It is with
this assumption that one can view a C60 molecule passing through a
double slit showing the familiar diffraction pattern as a single
quantum particle, its size notwithstanding \cite{Arndt}}.

Another important issue is the demarkation between microscopic,
mesoscopic and macroscopic.  Assuming that the object is made up of
$N$  physically relevant quantum particles (e.g. atoms, forgetting
about the tighter-bound substructures), starting with $N=1$  which we
refer to as microscopic, the question is: At what number of $N$  will
one begin to describe the object as mesoscopic with a qualitatively
distinct behavior from microscopic, and likewise for macroscopic? In
classical statistical mechanics, this issue underlies the important
attempt to derive from molecular Hamiltonian dynamics (with
deterministic chaos) the thermodynamic and kinetic properties of a
gas (with $N$  molecules), such as transport functions, and their
dynamics,  which possesses salient dissipative and time-asymmetric
features.  To provide a quantitative analysis of such issues one
needs to work with a stochastic equation for the $N$  particles so as
to be able to see the cross-over behavior between any two of these
three regimes as one varies $N$,  and its thermodynamic behavior as
one varies the temperature $T$.  In this paper we derive such an
equation for NHO in HOB, but will leave the analysis of this
theoretical issue to later investigations \cite{YCH}.

On the applied side, the master equation for NHO in a HOB is
useful for a range of problems which are of experimental
interests. Detection of small displacements of a NEM resonator by
a superconducting single electron transistor (SSET) or a biased
quantum point contact (BQPC) \cite{BlencoweRev} is a useful scheme
for probing the appearance of quantum properties of oscillators,
such as decoherence and entanglement \cite{ArmBleSch02}, noise and
fluctuations \cite{MMH04}, the standard quantum limit
\cite{SchSci04} and the uncertainty principle for quantum open
systems at finite temperatures \cite{HuZhaUncer,AndHal}. It has
been shown that in certain well-accessible regimes  both the SSET
and BQPC \cite{BIA05,HMM03} detection devices behave like a
thermal bath (albeit in some regimes it shows strong back-action
effects which have been suggested as a creative way to cool the
resonator \cite{ClerkBen05,SchNat06}). Master and Fokker-Planck
equations have been derived for these systems in
\cite{BIA05,RodArm}.

Another interesting setup is a linear array of NEM resonators.
Studying the entanglement transport in chains of mechanical
oscillators, Eisert {\it et. al.} \cite{Eisert} showed that the motions of
distant oscillators can be entangled without the need for control
of individual oscillators and without assuming any direct
interaction between them. Our master equation for $N$ harmonic
oscillators in a general environment can be usefully applied to
this problem to check on the distinct robustness of entanglement
in the canonical coordinates found by these authors.

Technically, our derivation of the master equation for the NHO system is
similar to the simple method we used in \cite{2HOQBM} for the case
of two coupled harmonic oscillators in a HOB. The task rests on finding a
suitable canonical transformation which preserves the structure of
the Poisson brackets. In Sec. 2 we define our model and outline
the procedure. In Sec. 3 we write down the evolution of the
density matrix and derive the master equation which has the same
form as the Hu-Paz-Zhang (HPZ) equation \cite{HPZ1,HPZ2,HY,StrunzYu2004}.
In Sec. 4,  we derive the corresponding Fokker-Planck equation. In Sec. 5 we
discuss the general features of these equations.  The overall
characteristics of the results derived in this paper may
posteriori be deduced from intuitive reasoning. However, the fine
points present in the full solution are not so easily obtained.
They are needed to address the validity of the COM axiom we posed
above, i.e., under what conditions can one presume that the center
of mass coordinate of a macroscopic object is the one most
sensitive to the environmental influence? We discuss this issue in
the last section. In the Appendices we derive the canonical
transformation for a general NHO system, and show the explicit
construction procedures for the cases of $N=2-5$.

\section{The Model}

For the investigation of environmental influences on the quantum
properties of a macroscopic object we consider the quantum Browninan
motion (QBM) of $N$ identical harmonic oscillators with mutual
interactions coupled to a collection of $n$ oscillators making up the
environment \cite{QBM1}. The generalized QBM Hamiltonian is:
\begin{eqnarray}
H_{\rm sys}= \sum_{i=1}^N (\frac{1}{2} M \dot{x}_i^2 + \frac{1}{2} M
\Omega^2 x_i^2) + \sum_{i,j = 1, i \neq j}^N V_{ij}(x_i-x_j)
\end{eqnarray}
\begin{eqnarray}
H_{\rm bath}= \sum_{i=1}^n (\frac{1}{2} m \dot{q}_i^2 + \frac{1}{2} m
\omega_i^2 q_i^2)
\end{eqnarray}
\begin{eqnarray}
H_{\rm int}= \sum_{i=1}^N \sum_{j=1}^n C_{ij}(x_i q_j)
\end{eqnarray}
For simplicity, let us assume that $C_{ij} = C_j, \forall i=1,...,N$
i.e. all the system harmonic oscillators couple to the bath with
equal strength. Hence the interacting Hamiltonian can be rewritten
more compactly as
\begin{eqnarray}
H_{\rm int}= \sum_{i=1}^N \sum_{j=1}^n C_{ij}(x_i q_j)= ( \sum_{i=1}^N
x_i ) ( \sum_{j=1}^n C_j q_j)
\end{eqnarray}
With this generalized $N$ harmonic oscillator (NHO) QBM model, we can
show the following:

(1) For any given finite $N$ sets of canonical coordinates $(x_i,
P_i), i=1,...,N$, we will give a procedure (or algorithm) to
construct another set of coordinates $(\tilde{X}_i, \tilde{P}_i),
i=1,...,N$ where $\tilde{X}_1 = \frac{1}{N}(x_1 +...+ x_N) ,
\tilde{P}_1 = P_1 + ... + P_N$ are the center-of-mass coordinate and
total momentum of the system respectively. This transformation matrix
$\hat{T}$ which transforms $\vec{x}$ to $\vec{\tilde{X}}$, i.e.
$\tilde{X}_i= \hat{T}_{ij}x_j$, satisfies $| \det{\hat{T}}| = 1$. The
choice of $\{ \tilde{X}_i \}$ is not unique. However, our method
gives a systematic way of construction and is thus quite useful. We
also provide the transformation for the effective masses. The
explicit construction is shown in Appendix A.

(2) The quadratic part of the system Hamiltonian can be shown to
transform as
\begin{eqnarray}
\sum_{i=1}^N (\frac{P_i^2}{2M} + \frac{1}{2}M\Omega^2 x_i^2) =
\sum_{i=1}^N (\frac{\tilde{P}_i^2}{2\tilde{M}_i} +
\frac{1}{2}\tilde{M}_i\Omega^2 \tilde{X}_i^2)
\end{eqnarray}
where $\tilde{M}_1 = \sum_{i=1}^N M = NM$ is the total mass of the
system.

Under this transformation, the Poisson brackets and hence the
commutation relations are preserved,
\begin{eqnarray}
[x_i, P_j]= [\tilde{X}_i, \tilde{P}_j]= i\hbar \delta_{ij},\quad
[x_i, x_j]=[P_i,P_j]=[\tilde{X}_i, \tilde{X}_j]=[\tilde{P}_i,
\tilde{P}_j]=0.
\end{eqnarray}
Hence our transformation is a canonical one. Most importantly, we
will show that the interaction potential between the N harmonic
oscillators in the new coordinate system is independent of
$\tilde{X}_1$,  the center-of-mass coordinate.\\

{\bf Lemma:} If the potentials $V_{ij}$ among the system
oscillators $x_i$ and $x_j$  are functions of $x_i - x_j$ only,
then $\frac{\partial V_{ij}}{\partial \tilde{X}_1} = 0$. Hence the
total potential $\sum_{i,j; i \neq j}^N V_{ij}(x_i - x_j)$ is
independent of $\tilde{X}_1$. The proof of the lemma will be given
in  Appendix B.

Combining all the above properties, we can rewrite the original
Hamiltonian as follows:
\begin{eqnarray}
H_{\rm sys} &=& \sum_{i=1}^N (\frac{1}{2} M \dot{x}_i^2 + \frac{1}{2}
M
\Omega^2 x_i^2) + \sum_{i,j = 1, i \neq j}^N V_{ij}(x_i-x_j) \\
&=& \sum_{i=1}^N (\frac{\tilde{P}_i^2}{2\tilde{M}_i} +
\frac{1}{2}\tilde{M}_i\Omega^2 \tilde{X}_i^2) +
\tilde{V}(\tilde{X}_2,...,\tilde{X}_N) \\ &=&
\frac{\tilde{P}_1^2}{2\tilde{M}_1} +
\frac{1}{2}\tilde{M}_1\Omega^2 \tilde{X}_1^2 + \sum_{i=2}^N
(\frac{\tilde{P}_i^2}{2\tilde{M}_i} +
\frac{1}{2}\tilde{M}_i\Omega^2 \tilde{X}_i^2) +
\tilde{V}(\tilde{X}_2,...,\tilde{X}_N) \\ &=& \tilde{H}_1 +
\tilde{H}_2,
\end{eqnarray}
where
\begin{eqnarray}
\tilde{H}_1 = \frac{\tilde{P}_1^2}{2\tilde{M}_1} +
\frac{1}{2}\tilde{M}_1\Omega^2 \tilde{X}_1^2 ,
\end{eqnarray}
\begin{eqnarray}
\tilde{H}_2 = \sum_{i=2}^N (\frac{\tilde{P}_i^2}{2\tilde{M}_i} +
\frac{1}{2}\tilde{M}_i\Omega^2 \tilde{X}_i^2) +
\tilde{V}(\tilde{X}_2,...,\tilde{X}_N).
\end{eqnarray}
\begin{eqnarray}
H_{\rm int}= \sum_{i=1}^N \sum_{j=1}^n C_{ij}(x_i q_j)= ( \sum_{i=1}^N
x_i ) ( \sum_{j=1}^n C_j q_j) =  \tilde{X}_1 (N \sum_{j=1}^n C_j
q_j)
\end{eqnarray}
Note that the heat bath has the spectral density $J(\omega)$:
\begin{eqnarray}
J(\omega)  = \pi  \sum_{j=1}^{n} \frac{\tilde{ C_j}^2}{2m_j
\omega_j} \delta(\omega - \omega_j).
\end{eqnarray} which differs from the original heat bath density by a numerical factor $N^2$.

The total Hamiltonian can then be written as
\begin{eqnarray*}
H_{\rm tot} &=& H_{\rm sys} + H_{\rm bath} + H_{\rm int} \\
&=& \tilde{H}_1 + \tilde{H}_2 + H_{\rm bath} + H_{\rm int} \\
& =& H_{\rm cm} + H^{\prime} + H_{\rm bath}.
\end{eqnarray*}
$H_{\rm cm} = \tilde{H}_1 + H_{\rm int}, H^{\prime} = \tilde{H}_2$. Note
that $ [H^{\prime},H_{cm} ]=[H^{\prime}, H_{\rm bath}]=0$.

\section{Density matrix and Master Equation} In our derivation, we
shall make the following two assumptions: (1) The system and the
environment are initially uncorrelated. (2) The heat bath is
initially in a thermal equilibrium state at temperature
$T=(k_B\beta)^{-1}$.

\subsection{The density matrix}
The density matrix for the total system develops in time under the
unitary evolutionary operator:
\begin{eqnarray}
\label{evolution1}
 \rho(t) &=& \exp\left[{-i\frac{H_{\rm tot}t}{\hbar}} \right]\rho(0)
 \exp\left[{i\frac{H_{\rm tot}t}{\hbar}}\right] \nonumber \\
 &=& \exp\left[{-i\frac{(H_{\rm cm}+H^{\prime} + H_{\rm bath})t}{\hbar}} \right]\rho(0)
 \exp\left[{i\frac{(H_{\rm cm}+H^{\prime} + H_{\rm bath})t}{\hbar}}\right] \nonumber \\
 &=&  \exp\left[{-i\frac{H^{\prime}t}{\hbar}} \right]
 \exp\left[{-i\frac{(H_{\rm cm} + H_{\rm bath})t}{\hbar}} \right]\rho(0)
 \exp\left[{i\frac{(H_{\rm cm} + H_{\rm bath})t}{\hbar}}\right]
  \exp\left[{i\frac{H^{\prime}t}{\hbar}}\right]
\end{eqnarray}
In the third equality, we have to use the condition that $[H_{\rm
cm}, H^{\prime}] = [H_{\rm bath}, H^{\prime}]=0$. (Note that for
general operators $A,B$, we have the Baker-Campbell-Hausdorff
formula: $e^A e^B = e^{A+B + \frac{1}{2}[A,B] +
\frac{1}{12}([A,[A,B]] + [B,[B,A]]) +...} $.)

If we define
\begin{eqnarray}
\label{evolution2}
 \tilde \rho(t) =  \exp\left[{-i\frac{(H_{\rm cm} + H_{\rm bath})t}{\hbar}} \right] \rho(0)
\exp\left[{i\frac{(H_{\rm cm} + H_{\rm bath})t}{\hbar}}\right]
\end{eqnarray}
then
\begin{eqnarray}
\label{evolution3}
 \rho(t) = \exp\left[{-i\frac{H^{\prime} t}{\hbar}}\right] \tilde \rho(t) \exp{\left[i\frac{H^{\prime}t}{\hbar}\right]}
\end{eqnarray}
\begin{eqnarray}
\rho_r=Tr_{\rm bath}\rho(t) =  \exp\left[{-i\frac{H^{\prime}
t}{\hbar}}\right] Tr_{\rm bath}\tilde\rho(t)
\exp{\left[i\frac{H^{\prime}t}{\hbar}\right]}
\end{eqnarray}
In a similar manner as in \cite{2HOQBM}, we can get the exact
master equation for the arbitrary $N$ oscillators.

\subsection{The Master equation}
Tracing over the heat bath leads us to a HPZ type master equation
\cite{HPZ1} for $\tilde{X}_1,\tilde{P}_1$:
\begin{equation}\label{masterosci}
\dot{\tilde\rho}_r = \frac{1}{i\hbar}[\tilde{H}_1,\tilde\rho_r]
+\frac{a(t)}{2i\hbar}[\tilde{X}_1^2,\tilde\rho_r] +
\frac{b(t)}{2i\hbar}[\tilde{X}_1,\{\tilde{P}_1,\tilde\rho_r\}]
+\frac{c(t)}{\hbar^2}[\tilde{X}_1,[\tilde{P}_1,\tilde\rho_r]] -
\frac{d(t)}{\hbar^2}[\tilde{X}_1,[\tilde{X}_1,\tilde\rho_r]]
\end{equation}
where $\tilde{H}_1$ is the Hamiltonian for
$\tilde{X}_1,\tilde{P}_1$ only. This has the same form as the HPZ
equation for the QBM of a single harmonic oscillator
$\tilde{X}_1,\tilde{P}_1$ interacting with a general heat bath.
Note that coefficients satisfy the same equations as listed in
Refs \cite{HPZ1} or \cite{HY}, but with
different coupling constants and masses.\\

From the  evolution equation (\ref{evolution3}), the required
master equation for $\rho_r(t)$ is obtained,
\begin{equation}\label{masterosci1}
\dot\rho_r = \frac{1}{i\hbar}[H_{\rm sys},\rho_r]
+\frac{a(t)}{2i\hbar}[\tilde{X}_1^2,\rho_r] +
\frac{b(t)}{2i\hbar}[\tilde{X}_1 ,\{\tilde{P}_1 ,\rho_r\}]
+\frac{c(t)}{\hbar^2}[\tilde{X}_1 , [\tilde{P}_1 ,\rho_r]] -
\frac{d(t)}{\hbar^2}[\tilde{X}_1 , [\tilde{X}_1 ,\rho_r]]
\end{equation}
The only difference between Eq. (\ref{masterosci1}) and Eq.
(\ref{masterosci}) is that the unitary evolution is modified by
the remaining $N-1$ fictitious harmonic oscillators $(\tilde{X}_j
, \tilde{P}_j), j=2,3,...,N$.

In terms of the original variables  $x_1,...,x_N, P_1,..., P_N$,
we get
\begin{eqnarray}
\label{mainresult}
 \dot\rho_r &=& \frac{1}{i\hbar}[H_{\rm sys},\rho_r]
+\frac{a(t)}{2 N^2i\hbar}[(x_1+...+x_N)^2,\rho_r] + \frac{b(t)}{2 N i\hbar}[x_1+...+x_N,\{P_1+...+P_N,\rho_r\}]\nonumber \\
&&+\frac{c(t)}{N \hbar^2}[x_1+...+x_N,[P_1+...+P_N,\rho_r]] -
\frac{d(t)}{N^2 \hbar^2}[x_1+...+x_N, [x_1+...+x_N,\rho_r]]
\end{eqnarray}
This exact master equation for the $N$ coupled harmonic
oscillators in a general environment is the main result of this
paper.

\section{Fokker-Planck equation}
In this section,  we present an alternative, but useful form
of the master equation derived in the last section.  We also
provide the explicit expressions for the coefficients appearing
in (\ref{masterosci1}).
\subsection{Fokker-Planck equation}
In terms of the Wigner function, the above master equation takes
the form:
\begin{eqnarray}
\label{fp} {\partial\tilde W \over\partial t}= & -&\sum_{i=1}^N
\left({P_i\over M}{\partial\tilde W\over \partial x_i} -
M\Omega^2 x_i{\partial\tilde W \over \partial P_i}\right)\nonumber \\
&+&M\Omega(t)(x_1+...+x_N)\left(\frac{\partial}{\partial
P_1}+...+\frac{\partial}{\partial P_N}\right)\tilde W +2
\Gamma(t)\left(\frac{\partial}{\partial P_1}
+...+\frac{\partial}{\partial P_N}\right)[(P_1 +...+ P_N)\tilde W] \nonumber \\
&+&\Sigma(t)\left(\frac{\partial}{\partial
P_1}+...+\frac{\partial}{\partial P_N}\right)^2\tilde W +
\Delta(t)\left(\frac{\partial}{\partial
P_1}+...+\frac{\partial}{\partial
P_N}\right)\left(\frac{\partial}{\partial
x_1}+...+\frac{\partial}{\partial x_N}\right)\tilde W
\end{eqnarray}
Note that the Wigner function is related to the reduced density
matrix in the following way:
\beqa
\label{wignerdensity} && \tilde W(x_1,..,x_N,P_1,..,P_N,t)\\
&&={1\over
{(2\pi)^N}}\int du_1..du_N\ e^{{i(u_1 P_1+...+u_NP_N)
/\hbar}}\rho_r\left(x_1-{u_1\over 2},..,x_N-{u_N\over
2};x_1+{u_1\over 2},..,x_N+{u_N\over 2},t\right)\>\nonumber
\eeqa
where we identify in Eq.~(\ref{fp}): \beqa
a(t)&=&M\Delta\Omega(t),\\
b(t)&=& 2 \Gamma(t),\\
c(t)&=&\Delta(t),\\
d(t)&=&\Sigma(t) \eeqa

By using this transformation, the master equation can be easily
obtained:
\begin{eqnarray}
i\hbar \frac{\partial \rho_r}{\partial t}  &=&
-\frac{\hbar^2}{2M}\left(\frac{\partial^2}{\partial x_1^2}  +...+
\frac{\partial^2}{\partial x_N^2}  - \frac{\partial^2}{\partial
y_1^2} -...- \frac{\partial^2}{\partial y_N^2} \right) \rho_r +
\frac{1}{2}M\Omega^2
(x_1^2  +...+ x_N^2 - y_1^2 -...- y_N^2 ) \rho_r \nonumber \\
&& + \frac{1}{2}M\delta \Omega^2(t)(x_1  +...+ x_N - y_1 -...-
y_N)\frac{1}{2}(x_1  +...+ x_N + y_1 +...+
y_N)\rho_r  \nonumber \\
&& -i \hbar \Gamma(t)(x_1 +...+ x_N - y_1 -...- y_N)\frac{1}{2}(
\frac{\partial}{\partial x_1} +...+ \frac{\partial}{\partial x_N}
- \frac{\partial}{\partial y_1} -...-
\frac{\partial}{\partial y_N} )\rho_r  \nonumber \\
&& -i M \Sigma(t)(x_1 +...+ x_N - y_1 -...- y_N)^2 \rho_r \nonumber  \\
&& + \hbar \Delta(t)(x_1 +...+ x_N - y_1 -...- y_N)(
\frac{\partial}{\partial x_1} +...+ \frac{\partial}{\partial x_N}
+ \frac{\partial}{\partial y_1} +...+ \frac{\partial}{\partial
y_N} )\rho_r
\end{eqnarray}

\subsection{Coefficients}
The coefficients $a(t), b(t), c(t), d(t)$ appearing in (\ref{mainresult}) or (\ref{fp}) can be constructed in terms of the elementary functions $u_i(s), i=1,2$. The $u_i(s)$
are defined as the functions that satisfy the following
homogeneous integro-differential equation
\begin{equation}{\ddot\Sigma}(s)+\Omega^2\Sigma(s)+{N^2\over M}\int_0
^s\,d\lambda\eta(s-\lambda)\Sigma(\lambda)=0
\end{equation}
with  the boundary conditions:
\begin{equation}u_1(s=0)=1\>,\,\,  u_1(s=t)=0\> ,
\end{equation}
and
\begin{equation}u_2(s=0)=0\>,\,\,  u_2(s=t)=1 \>.
\end{equation}
where
\begin{equation}
\eta(s)= -\int_0^{\infty} d\omega I(\omega) \sin(\omega s)
\end{equation}
is the dissipation kernel and $I(\omega)=\frac{1}{N^2} J(\omega)$ is the spectral density
of the environment. Note that the numerical pre-factor before the
integration in this equation is different from that defined in HPZ
\cite{HPZ1}. This is the main difference induced by the presence
of $N$ harmonic oscillators.

Let $G_1(s,\tau)$ be the Green function which satisfies the
following equation:
\begin{equation}
\frac{d^2}{ds^2}G_1(s,\tau) + \Omega^2 G_1(s,\tau) +
\frac{N^2}{M}\int_0^s d \tau \eta(s-\tau)G_1(s,\tau) =
\delta(s-\tau),
\end{equation}
where $G_1(s,\tau)$ as a function of $s$ satisfies the following
initial conditions:
\begin{equation}
G_1(s=0,\tau)= 0\>,\> \> {d\over ds}G_1(s,\tau)|_{s=0}=0\>.
\end{equation}
The Green function $G_2(s, \tau)$ is defined analogously. The
coefficients can then be written as
\begin{eqnarray}
 a(t) &=& N\int^t_0ds\eta(t-s)\left(u_2(s)-{u_1(s) \dot u_2(t)\over \dot
 u_1(t)}\right)\>,\\
 b(t) &=& {N\over M}\int^t_0ds\eta(t-s)\frac{u_1(s)}{\dot u_1(t)}\>.
\end{eqnarray}
\begin{eqnarray}
c(t)&=&{\hbar\over NM}\int^t_0\,d\lambda
G_1(t,\lambda)\nu(t-\lambda)\nonumber
\\
    & & -{N^2\hbar\over M^2}\int^t_0ds\int^t_sd\tau \int^t_0d\lambda
       \eta(t-s)G_1(t,\lambda)G_2(s,\tau)\nu(\tau-\lambda)\>,
\end{eqnarray}
and
\begin{eqnarray}
d(t)&=&\frac{\hbar}{N}\int^t_0\,d\lambda G'_1(t,\lambda)\nu(t-\lambda)\nonumber \\
    & & -{N^2\hbar\over M}\int^t_0ds\int^t_sd\tau \int^t_0d\lambda
        \eta(t-s)G'_1(t,\lambda)G_2(s,\tau)\nu(\tau-\lambda)\>.
\end{eqnarray}
where $\nu(s)$ is defined as
\begin{equation}
\nu(s)=\int^{+\infty}_0d\omega I(\omega)\coth({1\over
2}\hbar\omega\beta)\cos(\omega s)\>.
\end{equation}
which is the noise kernel of the environment. Here  a  prime
denotes taking the derivative with respect to the first variable
of $G_1(s,\tau)$.

\section{Discussions}

We end  with a few technical remarks followed by two conceptual
points, one referring to the COM axiom for the quantum dynamics of
macroscopic objects, and the other, to the generalized standard
quantum limit.

\subsection{Technical Remarks}
First, note that although we can show $[\tilde{X}_1,
\tilde{H}_2]=[\tilde{P}_1, \tilde{H}_2]=0$ and hence
$[\tilde{H}_1, \tilde{H}_2]=0$, the $(N-1)$ set of variables
$\tilde{X}_j, j=2,...,N$ will in general not commute with
$\tilde{H}_2$ because of the potential
$\tilde{V}(\tilde{X}_2,...,\tilde{X}_N)$.

Here we assume that all the $N$ particles are of the same mass $M$
and have the same eigenfrequency $\Omega$. They both couple to the
environment with equal strength. So these $N$ harmonic oscillators
are "identical" particles. If the N  system oscillators have
different masses, this becomes more involved.

In the proof that the potential is independent of $\tilde{X}_1$, we
made no assumption about the functional form of the potential. All
that was needed was that the potentials are functions of $x_i - x_j$
only. This is a reasonable assumption valid for many interesting
physical situations. Hence the range of applicability of our result
is by no means overly restrictive.

\subsection{The COM Axiom for quantum dynamics of macroscopic objects}

We now address the question raised in the beginning, i.e., on the
validity of representing the quantum behavior of a macroscopic
object by its center-of-mass dynamics, which we referred to as the
`COM axiom' for quantum dynamics of many body systems. To do this
we consider a more general type of coupling between the system and
the environment, e.g., coupling of the form $f(x_i)q_j$ instead of
$x_i q_j$, and examine if the COM variable dynamics separates from
the reduced variable dynamics.

For this purpose,  let us note that if the function $f(x)$ has the
property $\sum_{i=1}^{N} f(x_i) = \tilde{f}(\tilde{X}_1) +
g(\tilde{X}_2,...,\tilde{X}_N)$, for example $f(x) = x$ or $f(x) =
x^2$, one can split the coupling between the system and
environment into couplings containing the COM coordinate and the
relative coordinates. Tracing out the environmental degrees of
freedom $q_i$, one can easily get the influence action which
characterizes the effect of the environment on the system.

However, the coarse-graining made by tracing out the environmental
variables $q_i$ does not necessarily lead to the separation of the
COM and the relative coordinates in the effective action. When
they are mixed up and can no longer be written as the sum of these
two contributions, the form of the master equation will be
radically altered as it would contain both the relative coordinate
and the center-of-mass coordinate dynamics. One can work out how
much of a change this would bring about in the COM dynamics, but
at least we could say that when the coupling is not in these forms
the COM axiom for macroscopic quantum dynamics no longer holds.

Therefore we can conclude that for the N harmonic oscillators QBM
model, the coupling between the system and the environment need be
bi-linear, in the form $x_i q_j$,  for this axiom to hold. In that
case, one can say that the quantum evolution of a macroscopic
object in a general environment is completely described by the
dynamics of the center-of-mass variable obeying a master equation
of the HPZ type.

\subsection{Generalized standard quantum limit}

Another important issue of great interest to experimentalists is
the generalized uncertainty relation of an N-body system at finite
temperature. In our simple model the coupling between the system
of NHO and the environment is only through the center-of-mass
coordinate. If there is no mutual interaction between these N
harmonic oscillators, only the center-of-mass coordinate is
coupled to the environment and the remaining $(N-1)$ degrees of
freedom are orthogonal. Hence
 the uncertainty function  for the whole
system will simply be $U_{HZ}\times (U_{1})^{N-1}$, where $U_{HZ}$
is the uncertainty function for the 1HO QBM case given in
\cite{HuZhaUncer} and $U_{1}$  represents the quantum Heisenberg
uncertainty relation for one pair of relative coordinate canonical
variables $(\tilde{X}_i, \tilde{P}_i)$. Here we want to point out
that in this situation, the effect of the environment, say the
temperature of the heat bath, enters only through  the
center-of-mass coordinate in the form of $U_{HZ}$, the remaining
$(N-1)$ pair of canonical conjugate variables are subject to
quantum Heisenberg uncertainty relation $U_{1}$. The time
evolution and temperature dependence of the uncertainty function
for an Ohmic bath was studied in great detail in
\cite{HuZhaUncer,AndHal}.

If inter-particle interactions $V_{ij} \neq 0 $ exist among the $N$
particles, then the uncertainty function $U_{HZ}\times (U_{1})^N$
will be modified. Due to the interactions between the $N$ harmonic
oscillators, the uncertainty relation governing the $(N-1)$ pairs of
relative coordinate variables $(\tilde{X}_i, \tilde{P}_i)$ might be
squeezed and rotated. However,  if the  number of harmonic
oscillators $N$ is large, and the interacting potential among the
N-particles are short-ranged, or that the forces amongst them are
very strong, so that the characteristic frequencies of these `hard
modes' are much higher than that of the natural frequency of the COM
modes, under these conditions, the quanta corresponding to the motion
of the relative coordinates are not easily excited and the leading
order contribution to the uncertainty function will be dominated by
the center-of-mass degree of freedom. The details can be worked out
from a perturbation analysis on the present results. (See, e.g.,
\cite{HPZ2} for treating one form of interaction.)

\subsection{Conclusion}

In this paper, we outlined the procedure to find a canonical
transformation to transform from the individual coordinates $(
x_i, P_i )$ to the collective coordinates $( \tilde{X}_i,
\tilde{P}_i ), i=1,...,N$ where $\tilde{X}_1, \tilde{P}_1$ are the
center-of-mass coordinate and momentum respectively. We then
proved that the potential $V_{ij}(x_i - x_j)$ is independent of
the center-of-mass coordinate $\tilde{X}_1$. Then following the
simple derivation of the master equation for 2HO in  our previous
work we showed that the system with variables
($\tilde{X}_1,\tilde{P}_1$) obeys a master equation of the same
form as the HPZ equation. We gave the details of derivation of an
exact non-Markovian master equation for the reduced density matrix
of this system constructed with the heat bath variables integrated
out. We stress that this result for the $N$ mutually interacting
harmonic oscillators in a general environment is more than just a
normal mode decomposition problem as in classical mechanics
because there are interactions between the individual system
oscillators and collective interaction with the quantum
environment.

This result is expected to be useful for the study of entanglement
dynamics of multipartite particles and  quantum to classical
transition issues. Finally,  we established a relation between the
center-of-mass coordinate of $N$ body harmonic oscillators and the
well-known one oscillator QBM model. This provides a key step in
establishing a microscopic theory for macroscopic quantum phenomena,
a topic we intend to pursue further in the future.

\section{Appendix}

\subsection{Construction of the canonical transformation for general $N$. }
Given a system of $N$ identical harmonic oscillators with equal
mass $M$ and intrinsic frequency $\Omega$.
\begin{eqnarray*}
H_{N0} = \sum_i^N \frac{P^2_i}{2M }  +  \sum_i^N \frac{1}{2} M
\Omega^2 x_i^2 ,
\end{eqnarray*}
Note that $P_i = M \dot{x}_i = \frac{\partial H_{N0}}{\partial
\dot{x}_i}$ and $[x_i, P_j] = i \hbar \delta_{ij}$.

Goal: Find out $\{ \tilde{X}_i  \}$, $\{ \tilde{P}_i  \}$, and $\{
\tilde{M}_i \}$ such that
\begin{eqnarray*}
H_{N0} = \sum_i^N \frac{P^2_i}{2M }  +  \sum_i^N \frac{1}{2} M
\Omega^2 x_i^2 = \sum_i^N \frac{\tilde{P}^2_i}{2 \tilde{M}_i } +
\sum_i^N \frac{1}{2} \tilde{M}_i \Omega^2 \tilde{X}_i^2 ,
\end{eqnarray*}
where $[\tilde{X}_i, \tilde{P}_j] = i \hbar \delta_{ij} = [x_i,
P_j]$ and $\tilde{X}_1=\frac{1}{N}(x_1 + x_2 + ...+ x_N),
\tilde{P}_1 = (P_1 + P_2 + ...+ P_N)$ and $\tilde{M}_1 = NM$ are
the center-of-mass coordinate,  total momentum, and total mass of
the system respectively. Note that it's easy to see that
$[\tilde{X}_1,\tilde{P}_1]=i\hbar$.\\

\subsection{The explicit construction procedure:}
\subsubsection{$N = 2$:}
\begin{eqnarray*}
\tilde{X}_1&=&\frac{1}{2}(x_1 + x_2), \quad \tilde{X}_2 = x_1 -
x_2,\\ \tilde{P}_1 &=& P_1 + P_2, \quad
\tilde{P}_2=\frac{1}{2}(P_1-P_2), \quad [\tilde{X}_i,\tilde{P}_j]
= i \hbar \delta_{ij}.
\end{eqnarray*}
\begin{eqnarray*}
(\tilde{M}_1, \tilde{M}_2) = (2M, \frac{M}{2})
\end{eqnarray*}
Note that $\tilde{P}_i = \tilde{M}_i \dot{X}_i$ and $P_i = M
\dot{x}_i$.
\begin{eqnarray}
&& \frac{P_1^2}{2M} + \frac{P_2^2}{2M}  + \frac{1}{2}M \Omega^2
x_1^2 + \frac{1}{2}M \Omega^2 x_2^2
\\ &=& \frac{\tilde{P}_1^2}{2(2M)} +
\frac{\tilde{P}_2^2}{2(\frac{M}{2})}  + \frac{1}{2}(2M) \Omega^2
\tilde{X}_1^2 + \frac{1}{2}(\frac{M}{2}) \Omega^2 \tilde{X}_2^2
\end{eqnarray}
\subsubsection{$N = 3 $}:
We can go from canonical variables $(x_1, x_2, x_3)$ to
$(\tilde{X}_1=X_{cm}, \tilde{X}_2, \tilde{X}_3)$ with the
following transformation: Viewing $\{ x_i\}$ and $\{ \tilde{X}_i
\}$ as orthogonal base of the 3 dimensional vector space, our goal
is to find a $3 \times 3$   linear transformation matrix $T_{ij}$,
such that $\tilde{X}_i = T_{ij} x_j$ with $\det(T_{ij})=1$. A
convenient choice is that since we have already construct the
transformation of $N=2$ case, we can just choose  $\tilde{X}_1 =
\frac{1}{3}(x_1+ x_2+ x_3)$ and $\tilde{X}_2 = x_1 - x_2$. Since
$\tilde{X}_3$ must perpendicular with $\tilde{X}_1, \tilde{X}_2$,
we have $\tilde{X}_3 = a_3 (x_1 + x_2 - 2 x_3)$. Since
$\det(T)=1$,  we then have $a_3=\frac{1}{2}$.

Note that $\tilde{P}_3 = \tilde{M}_3 \dot{X}_3$ hence proportional
to $\tilde{X}_3$, thus we have $\tilde{P}_3 = b_3 (P_1 + P_2 - 2
P_3)$. From $[\tilde{X}_3,\tilde{P}_3]=i \hbar$, we have
$b_3=\frac{1}{3}$.
\begin{eqnarray*}
\tilde{X}_1 = \frac{1}{3}(x_1+ x_2+ x_3)= X_{cm}, \quad
\tilde{X}_2 = x_1 - x_2, \quad \tilde{X}_3 = \frac{1}{2}(x_1 + x_2
- 2x_3),
\end{eqnarray*}
\begin{eqnarray*}
\tilde{P}_1 = (P_1+ P_2+ P_3), \quad \tilde{P}_2 = \frac{1}{2}(P_1
- P_2), \quad \tilde{P}_3 = \frac{1}{3}(P_1 + P_2 - 2 P_3)
\end{eqnarray*}
and we have
\begin{eqnarray*}
[\tilde{X}_i, \tilde{P}_j ] = i \hbar \delta_{ij}, \quad
[\tilde{X}_i, \tilde{X}_j] = [\tilde{P}_i, \tilde{P}_j] = 0.
\end{eqnarray*}
\begin{eqnarray*}
(\tilde{M}_1, \tilde{M}_2, \tilde{M}_3) = (2M, \frac{M}{2},
\frac{2 M}{3}=\frac{b}{a}M)
\end{eqnarray*}
\begin{eqnarray*}
&& \frac{P_1^2}{2M} + \frac{P_2^2}{2M} + \frac{P_3^2}{2M} +
\frac{1}{2}M \Omega^2 x_1^2 + \frac{1}{2}M \Omega^2 x_2^2 +
\frac{1}{2}M \Omega^2 x_3^2 \\ &=& \frac{\tilde{P}_1^2}{2(3M)} +
\frac{\tilde{P}_2^2}{2(\frac{M}{2})} +
\frac{\tilde{P}_3^2}{2(\frac{2M}{3})} + \frac{1}{2}(3M) \Omega^2
\tilde{X}_1^2 + \frac{1}{2}(\frac{M}{2}) \Omega^2 \tilde{X}_2^2 +
\frac{1}{2}(\frac{2M}{3}) \Omega^2 \tilde{X}_3^2
\end{eqnarray*}

\subsubsection{ $N = 4$:}
Note that in this case N=4=$2 \times 2$. We can make use of our result of N=2.  \\
Since we require that $\tilde{X}_1 = \frac{1}{4}(x_1 + x_2 + x_3 +
x_4)$ in this case. We may first define $y_1 = \frac{1}{2}(x_1 +
x_2),$
 $y_2 = \frac{1}{2}(x_3 + x_4)$, and  $y_3 =(x_1 - x_2), y_4 =
(x_3 -x_4)$. That means that we treat the whole 4 oscillators as
two pairs of oscillators using the center-of-mass and relative
coordinates of each pair. The corresponding conjugate momentum are
$P_{y1}= (P_1 + P_2), P_{y2}=(P_3 + P_4), P_{y3}=\frac{1}{2}(P_1
-P_2), P_{y4} = \frac{1}{2}(P_3 - P_4)$. With this transformation,
we have $[y_i, P_{yj}] = i \hbar \delta_{ij}$. The effective
masses for $y_i$ would be $(2M, 2M, \frac{M}{2}, \frac{M}{2})$.
It's easy to see that $y_3$ is perpendicular to all the other
$y_i$, and so is $y_4$. Hence $y_3 , y_4$ are perpendicular to any
linear combination of $y_1$ and $y_2$. That means by changing from
$x_i$ to $y_i$ we can decompose the original 4 dimensional vector
space generated by $x_i$ into direct sums of a two dimensional
vector space generated by $y_1$ and $y_2$ and two one dimensional
vector spaces generated by $y_3$ and $y_4$ respectively.\\

We can then make use of $N = 2$ result for $y_1, y_2$. Hence
$\tilde{X}_1 = \frac{1}{2}(y_1 + y_2)  = \frac{1}{4}(x_1 + x_2 +
x_3 + x_4), \tilde{X}_2 = y_1 - y_2 = \frac{1}{2}(x_1 + x_2 - x_3
-x_4).$ The corresponding $\tilde{P}_1 = P_{y1} + P_{y2} = P_1 +
P_2 + P_3 + P_4, \tilde{P}_2 = \frac{1}{2}(P_{y1} -P_{y2}) =
\frac{1}{2}(P_1 + P_2 - P_3 - P_4)$. The effective masses are
$(2(2M), \frac{1}{2}(2M)) = (4M, M)$.\\

Combing with the other two one dimensional vector spaces generated
by $y_3$ and $y_4$, we then have the following transformation
\begin{eqnarray}
\tilde{X}_1 &=& \frac{1}{2}(y_1 + y_2)  = \frac{1}{4}(x_1 + x_2 +
x_3 + x_4), \nonumber \\ \tilde{X}_2 &=& y_1 - y_2 =
\frac{1}{2}(x_1 + x_2 - x_3
-x_4) \nonumber \\
\tilde{X}_3 &=& y_3 = x_1 - x_2, \nonumber \\ \tilde{X}_4 &=& y_4
= x_3 - x_4
\end{eqnarray}
with corresponding conjugate momentum
\begin{eqnarray}
\tilde{P}_1 &=& P_{y1} + P_{y2} = P_1 + P_2 + P_3 + P_4, \nonumber
\\ \tilde{P}_2 &=& \frac{1}{2}(P_{y1} -P_{y2}) = \frac{1}{2}(P_1 +
P_2 - P_3 - P_4),  \nonumber \\ \tilde{P}_3 &=&
P_{y3}=\frac{1}{2}(P_1 -P_2),  \nonumber \\ \tilde{P}_4 &=& P_{y4}
= \frac{1}{2}(P_3 - P_4)
\end{eqnarray}
and we have
\begin{eqnarray*}
[\tilde{X}_i, \tilde{P}_j ] = i \hbar \delta_{ij}, \quad
[\tilde{X}_i, \tilde{X}_j] = [\tilde{P}_i, \tilde{P}_j] = 0.
\end{eqnarray*}
\begin{eqnarray*}
(\tilde{M}_1, \tilde{M}_2, \tilde{M}_3, \tilde{M}_4) = (4M, M,
\frac{M}{2}, \frac{ M}{2})
\end{eqnarray*}
This procedure can be easily applied to any other cases where the
number of oscillators are even. For example if $N = 2 \times k$.
We can first define $y_1=\frac{1}{2}(x_1 + x_2),
y_2=\frac{1}{2}(x_3 + x_4), ... , y_k=\frac{1}{2}(x_{2k-1} +
x_{2k})$ and $y_{k+1} = (x_1 - x_2), y_{k+2} = (x_3 - x_4), ... ,
y_{2k} = (x_{2k-1} - x_{2k})$. The conjugate momentum are
$P_{y1}=(P_1 + P_2), P_{y2} = (P_3 + P_4), ... , P_{yk}= (P_{2k-1}
+ P_{2k}), P_{yk+1} = \frac{1}{2}(P_1 -P_2), P_{yk+2} =
\frac{1}{2}(P_3 -P_4), ..., P_{y2k} = \frac{1}{2}(P_{2k-1} -
P_{2k})$ and the corresponding effective masses are $(2M,
2M,...2M, \frac{M}{2}, \frac{M}{2},..., \frac{M}{2})$.\\

We then make use of the transformations for the $N = k$ case to
$\{ y_i  \}, i=1,...,k$. By doing this we can have $\{ \tilde{X}_i
\}, \{  \tilde{P}_i \}, \{ \tilde{M}_i\}, i=1,...,k$. Together
with the $\tilde{X}_j = y_{j}, \tilde{P}_j = P_{j}, j=k+1,...,2k$
, we then have the complete transformation for $N = 2k$ case.

\subsubsection{$N = 5 $:}
 Because $5=4+1$, we can make use of the result
for $N = 4$ case. But note that in our construction we require
that $\tilde{X}_1 = \frac{1}{5}(x_1 + x_2 + x_3 + x_4 + x_5)$ in
this case, hence the corresponding $\tilde{P}_1 = (P_1 + P_2 + P_3
+ P_4 + P_5)$ and effective mass is $5M$. The corresponding
$\tilde{X}_2, \tilde{X}_3, \tilde{X}_4$ and $\tilde{P}_2,
\tilde{P}_3, \tilde{P}_4$ and effective masses are the same as
those in the case of the $N = 4$ case. The remaining $\tilde{X}_5$
has to be perpendicular to $\tilde{X}_i , i=1,2,3,4$ and it's thus
obvious that it must be a linear combination of the
$\tilde{X}^{\prime}_1 = \frac{1}{4}(x_1 + x_2 + x_3 + x_4)$ in
$N=4$ case and $x_5$ such that it is perpendicular to $\tilde{X}_1
= \frac{1}{5}(x_1 + x_2 + x_3 + x_4 + x_5)$. It's easy to see that
$\tilde{X}_5 = a_5 (x_1 + x_2 + x_3 + x_4 - 4 x_5)$. Hence the
corresponding conjugate momentum is $\tilde{P}_5 = b_5 ( P_1 + P_2
+ P_3 + P_4 - 4 P_5 )$. With the requirement that $\tilde{X}_i =
T_{ij}x_j$ and $\det(T_{ij}) =1$, we have $a_5 = \frac{1}{4}$.
 From
the requirement $[\tilde{X}_5, \tilde{P}_5] = i \hbar$, we then
have $b_5 = \frac{1}{5}$. The corresponding effective mass is then
given by $\tilde{M}_5 = \frac{b_5}{a_5}M = \frac{4}{5}M$.\\

Hence the transformation for $N = 5$ is given by
\begin{eqnarray}
\tilde{X}_1 &=&  \frac{1}{5}(x_1 + x_2 + x_3 + x_4 + x_5), \nonumber \\
\tilde{X}_2 &=&  \frac{1}{2}(x_1 + x_2 - x_3
-x_4) \nonumber \\
\tilde{X}_3 &=&  x_1 - x_2, \nonumber \\ \tilde{X}_4 &=&  x_3 -
x_4,  \nonumber \\ \tilde{X}_5 &=& \frac{1}{4}(x_1 + x_2 + x_3 +
x_4 - 4 x_5),
\end{eqnarray}
with corresponding conjugate momentum
\begin{eqnarray}
\tilde{P}_1 &=&  P_1 + P_2 + P_3 + P_4 + P_5 , \nonumber
\\ \tilde{P}_2 &=&  \frac{1}{2}(P_1 +
P_2 - P_3 - P_4),  \nonumber \\ \tilde{P}_3 &=& \frac{1}{2}(P_1
-P_2),  \nonumber \\ \tilde{P}_4 &=&  \frac{1}{2}(P_3 - P_4),
\nonumber \\ \tilde{P}_5 &=& \frac{1}{5}(P_1 + P_2 + P_3 + P_4 -4
P_5),
\end{eqnarray}
and we have
\begin{eqnarray*}
[\tilde{X}_i, \tilde{P}_j ] = i \hbar \delta_{ij}, \quad
[\tilde{X}_i, \tilde{X}_j] = [\tilde{P}_i, \tilde{P}_j] = 0.
\end{eqnarray*}
\begin{eqnarray*}
(\tilde{M}_1, \tilde{M}_2, \tilde{M}_3, \tilde{M}_4, \tilde{M}_5)
= (5M, M, \frac{M}{2}, \frac{ M}{2}, \frac{4M}{5}).
\end{eqnarray*}

This procedure can be easily applied to any other cases where the
number of oscillators are odd. For example if $N = 2  k +1$. We
can make use of the results for $N = 2 k$. Since we already have
$\{x_i \}$ to $\{ \tilde{X}_i\}, i=1,...,2k$ and these
$\tilde{X}_i$ are already orthogonal to each other. With the
addition of $x_{2k+1}$ we only need to change $\tilde{X}_1 =
\frac{1}{2k +1}(x_1 + x_2 + ... + x_{2k} + x_{2k+1})$ and it's
obvious that the $\tilde{X}_{2k+1}$ must have the structure
$\tilde{X}_{2k+1} = a_{2k+1}(x_1 + x_2 + ... + x_{2k} - 2k \,
x_{2k+1})$. With the requirement that $\tilde{X}_i = T_{ij}x_j$
and $\det(T_{ij}) =1$, we have $a_{2k+1} = \frac{1}{2k}$.
Similarly $\tilde{P}_{2k+1} = b_{2k+1}(P_1 + P_2 + ... + P_{2k}
-2k \, P_{2k+1})$. The requirement $[\tilde{X}_{2k+1},
\tilde{P}_{2k+1}]= i \hbar $ gives $b_{2k+1} = \frac{1}{2k+1}$
hence the corresponding effective mass is $\tilde{M}_{2k+1} =
\frac{b_{2k+1}}{a_{2k+1}}M = \frac{2k M}{2k+1}=\frac{N-1}{N}M$.

We have thus provided a procedure to explicitly construct the
canonical transformations  for the canonical coordinates for any
finite number of harmonic oscillators. The transformation for the
effective masses are also given. With this canonical
transformation, the structure of the Poisson brackets are
preserved.

\subsection{Proof of the Lemma}

{\bf Lemma:} If the potentials $V_{ij}$ among the system
oscillators $x_i$ and $x_j$  are functions of $x_i - x_j$ only,
then $\frac{\partial V_{ij}}{\partial \tilde{X}_1} = 0$. Hence the
total potential $\sum_{i,j; i \neq j}^N V_{ij}(x_i - x_j)$ is
independent of $\tilde{X}_1$.\\

{\bf Proof of the Lemma:} The reason is very simple: regarding the
original $\{ x_i \}, i=1,...,N$ as an orthonormal coordinate basis
which spans a $N$-dimensional vector space. My construction for
$\{ \tilde{X}_i \}, i=1,...,N$ is another set of complete
orthogonal basis which contains the center-of-mass coordinate
$X_{cm}=\frac{1}{N}(x_1+...+x_N)$ as $\tilde{X}_1$. In the $\{ x_i
\}$ basis, $\tilde{X}_1 = \frac{1}{N}(1,1,...,1)$ and $x_i-x_j =
(0,0,...,1,0,..,-1,0,..,0)$ with the i-th entry equals to 1 and
the j-th entry equals to -1. Hence it is obvious that $x_i - x_j$
is orthogonal to $X_{cm}= \tilde{X}_1$. Hence $x_i - x_j$ belongs
to the $(N-1)$-dimensional vector space spanned by $\{ \tilde{X}_j
\},j=2,...,N$ of the new coordinates. Because $\{ \tilde{X}_i \},
i=1,...,N$ is an orthogonal basis of the N-dimensional vector
space $V^N$, $\{ \tilde{X}_j \}, j=2,...,N$ is an orthogonal basis
for the (N-1)-dimensional subspace $W^{N-1}$. Let $L$ be the
1-dimensional vector space spanned by $\tilde{X}_1$, then $V^N = L
\oplus W^{N-1}, L\cap W^{N-1} =0 $. Hence $(x_i - x_j) \in
W^{N-1}$ and can be expressed in terms of $\{ \tilde{X}_j \},
j=2,...,N$ uniquely. Since we assume that the potentials are
functions of only $x_i-x_j$, hence the potentials are functions of
$\{ \tilde{X}_j \},j=2,...,N$ only and are thus independent of
$\tilde{X}_1$. \quad \quad \quad {\bf Q.E.D.}

\begin{acknowledgments}
The authors acknowledge support from the following funding
agencies: CHC by the National Science Council of Taiwan under
Grant Nos. NSC93-2112-M-006-011. BLH by the NSF (PHY-0426696)
under the ITR program and by NSA-LPS to the University of
Maryland.  TY by  ARO Grant W911NF-05-1-0543 to the University of
Rochester. Part of this work was done while we enjoyed the
hospitality of the Institute of Physics of the Academia Sinica,
Taipei, the National Center for Theoretical Sciences and the
Center for Quantum Information Sciences at the National Cheng Kung
University, Tainan, Taiwan.
\end{acknowledgments}


\end{document}